\documentclass{svjour3}                     
\smartqed  
\usepackage{graphicx}
\usepackage{amsmath, amssymb, amsfonts}
\begin{document}

\title{A new method of description of three-particle Coulombic systems}


\author{V.~B. Belyaev   \and
        A.~A. Naumkin   
}


\institute{V.~B. Belyaev \and A.~A. Naumkin \at
              BLTP JINR, 141980 Dubna, Moscow region, Russia \\
              Tel.: +7-49621-65-900\\
              Fax: +7-49621-65-084\\
              \email{belyaev@theor.jinr.ru}           
           \and
           A.~A. Naumkin \at
              Department of Physics, Moscow State University, 119992 Moscow, Russia \\
              \email{naumkin@physics.msu.ru}           
}

\date{Received: date / Accepted: date}

\maketitle

\begin{abstract}
We present a method for treatment of three charged particles. The
proposed method has universal character and is applicable both for
bound and continuum states. A finite rank approximation is used
for Coulomb potential in three-body system Hamiltonian, that
results in a system of one-dimensional coupled integral equations.
Preliminary numerical results for three-body atomic and molecular
systems like $H^{-}$, $He$, $pp\mu$ and other are presented.
\end{abstract}

\section{Introduction}
\label{intro}

The quantum three-body problem emerges in various fields of
physics, and different methods of treating it are developed.
However, there are no universal methods able to solve it in case
of charged particles in the continuum. These problem is important
in atomic and molecular physics and in nuclear astrophysics.

The main purpose of the work is to develop a procedure applicable
to treatment of three charged particles in continuum. One of the
ways to construct it is to make an approximation on operator
level, i.e. in the Hamiltonian of three-body system under
consideration. After that any boundary conditions can be used. The
idea of our work was inspired by earlier paper of A. Weinstein~\cite{Weinst_37},
who introduced the so-called method of intermediate
Hamiltonians. This method was applied to calculate lower bounds
for eigenvalues of some differential operators. N.~W. Bazley and
D.~W. Fox applied it to $He$ atom and other physical systems
\cite{BazFox_61,Bazley_60}. They constructed sequence of
intermediate Hamiltonians using finite rank operators. These
operators are defined in the whole space of full Hamiltonian. In
opposite to that we will apply finite rank approximation in a
subspace of the three-body system Hamiltonian, namely in the
angular space of hyperspherical variables. This results in a
system of coupled one-dimensional integral equations.

In the following sections we review a method of intermediate
Hamiltonians, 
derive system of integral equations and report the results of calculations.

\section{Hyperspherical coordinates}

There are different ways to formulate the three-body problem. We
use hyperspherical coordinates, and in this section we give a
brief review of them. Complete theory, derivations etc can be
found, e.g., in \cite{Lin_95}.

We start with the three-body system Hamiltonian:
\begin{equation}
\label{Hamilt} H = -\sum_{i=1}^{3} \frac{1}{2m_{i}}\nabla_{i}^{2}+
\sum_{i<j} V_{ij}({\bf r}_{i}-{\bf r}_{j}),
\end{equation}
where $i$ enumerates different particles and corresponding sets od
Jackobi coordinates.is a numb${\bf r}_{i}$ is a position vector of
the $i$-th particle. The scaled Jackobi coordinates are introduced
as follows:
\begin{equation}
\begin{array}{rcl}\cr
        {\bf x}_i&=&\displaystyle\left[\frac{m_{j} m_{k}}{m_{j}+m_{k}}\right]^{1/2}({\bf r}_{j}-{\bf r}_{k})\cr
        {\bf y}_i&=&\displaystyle\left[\frac {m_{i}(m_{j}+m_{k})} {m_{1}+m_{2}+m_{3}} \right]^{1/2}
 \left(-{\bf r}_{i}+\frac{m_{j}{\bf r}_{j}+m_{k}{\bf r}_{k}}{m_{j}+m_{k}}
 \right)
\end{array}
\label{Jacoord}
\end{equation}
and the Hamiltonian~(\ref{Hamilt}) takes the form
\begin{equation}
\label{Ham_jac} H = -\frac{1}{2}\nabla_{\bf x}^{2}
-\frac{1}{2}\nabla_{\bf y}^{2}+ V,
\end{equation}
where $V = V_{13}+V_{23}+V_{31}$ --- a sum of pair potentials.
Taking ${\bf x}$ and ${\bf y}$ in spherical coordinates $({\bf
x},{\bf y}) \rightarrow
(x,\theta_{1},\varphi_{1},y,\theta_{2},\varphi_{2})$, one obtains:
\begin{equation}
\label{Ham_jac_sp} \hat{H} =
-\frac{1}{2x^2}\frac{\partial}{\partial x}
\left(x^{2}\frac{\partial}{\partial x}\right) -
\frac{1}{2x^2}\Delta_{\Omega_1} -
\frac{1}{2x^2}\frac{\partial}{\partial x}
\left(x^{2}\frac{\partial}{\partial x}\right) -
\frac{1}{2y^2}\Delta_{\Omega_2}+ V,
\end{equation}
Now let us introduce hyperspherical variables:
\begin{equation}
\label{Hyp_coord} x = \rho \cos \alpha, \quad y = \rho \sin \alpha
\end{equation}
Here $\rho$ is hyperradius, $\alpha$ --- hyperangle. Hamiltonian
expressed in terms of this variables has the form:
\begin{equation}
\label{Ham_hyp} \hat{H} =
-\frac{1}{2}\left(\frac{\partial^2}{\partial \rho^2} +
\frac{5}{\rho}\frac{\partial}{\partial \rho}\right)
- \frac{1}{2\rho^2}\left[ \frac{\partial^2}{\partial \alpha^2} +
4\cot 2\alpha \frac{\partial}{\partial \alpha} +
\frac{1}{\cos^{2}\alpha}\Delta_{\Omega_1} +
\frac{1}{\sin^{2}\alpha}\Delta_{\Omega_2}\right] + V
\end{equation}
Angular part of a kinetic energy operator is the hypermomentum
operator:
\begin{equation}
\label{K_ang} \hat{K} = \frac{\partial^2}{\partial \alpha^2} +
4\cot 2\alpha \frac{\partial}{\partial \alpha} +
\frac{1}{\cos^{2}\alpha}\Delta_{\Omega_1} +
\frac{1}{\sin^{2}\alpha}\Delta_{\Omega_2},
\end{equation}
and its eigenfunctions are hyperspherical harmonics:
\begin{equation}
\label{HypFunc} {\mathcal Y}_{K}^{l_{1}m_{1}l_{2}m_{2}}(\alpha,
\Omega_1, \Omega_2) = c_{K}^{l_{1}l_{2}}
(\sin\alpha)^{l_{1}}(\cos\alpha)^{l_{2}}
P_{n}^{(l_{1}+\frac{1}{2},l_{2}+\frac{1}{2})}(\cos 2\alpha)
Y_{l_{1}m_{1}}(\Omega_1) Y_{l_{2}m_{2}}(\Omega_2),
\end{equation}
where
\begin{equation}
c_{K}^{l_{1}l_{2}} =
\left[\frac{2n!(K+2)(n+l_{1}+l_{2}+1)!}{\Gamma(n+l_{1}+3/2)\Gamma(n+l_{2}+3/2)}\right]^{1/2}.
\end{equation}
Let us consider system of three particles with masses
$m_{1},m_{2},m_{3}$ and charges $q_{1},q_{2},q_{3}$. The Coulomb
potential has the form:
\begin{equation}
\label{Pot_Coul_J} V({\bf x}, {\bf y}) =
\frac{b_1}{x_1}+\frac{b_2}{x_2}+\frac{b_3}{x_3},
\end{equation}
where $b_{i}=\sqrt{\frac{m_{j}m_{k}}{m_{j}+m_{k}}}q_{j}q_{k}$. In
hyperspherical coordinates:
\begin{equation}
\label{Pot_Coul_H} V(\rho,\Omega) =
\frac{1}{\rho}\left(\frac{b_1}{\cos\alpha_{1}}+\frac{b_2}{\cos\alpha_{2}}
+\frac{b_3}{\cos\alpha_{3}}\right)
\end{equation}
Here $\alpha_{i}$ --- hyperangles corresponding to different sets
of Jacobi coordinates.

\section{Finite rank operators}

Finite rank operators are widely used in different problems of
mathematical physics. They allow one to reduce complexity of a
problem and proceed to its solution. E.g., in \cite{BelWrz_79}
finite-rank operator was used to describe nuclear part of full
Hamiltonian in a problem of low energy $\pi - {^{3}He}$
scattering.

N.~W. Bazley and D.~W. Fox used finite rank operators to calculate
lower bounds of eigenvalues of Schr\"odinger eqution
\cite{BazFox_61,Bazley_60}. Let us shortly review the method of
intermediate Hamiltonians they used.

We suppose that full Hamiltonian $H$ can be presented as a sum of
$H^0$, that has known eigenvalues and eigenfunctions, and a
positively definite $H'$. The exactly solvable Hamiltonian $H^{0}$
is assumed to have ordered discrete energy levels $E_{1}^{0} \leq
E_{1}^{0} \leq ...$ below its continuum spectrum. The
corresponding eigenfunctions are $\psi_{1}^{0}$, and we have
\begin{equation}
H^{0} \psi_{i}^{0} = E_{i}^{0} \psi_{i}^{0}.
\end{equation}
Since $H = H^{0} + H'$, where $H'$ is positively definite, $H^{0}
\leq H$ and $E_{1}^{0} \leq E_{1}$. Thus, the full Hamiltonian $H$
and $H^{0}$ are linked by a sequence of intermediate Hamiltonians:
\begin{equation}
H^{0} \leq H^{k} \leq H^{k+1}  \leq H.
\end{equation}
To construct the Hamiltonians $H^{k}$, we introduce a system of
$k$ linearly independent functions $p_{1}, p_{2}, ..., p_{k}$. The
set of functions $p_{1}, p_{2}, ...$ is defined in the whole space
of definition of the Hamiltonian H. Projection of some
wavefunction $\varphi$ on these functions is given by
\begin{equation}
P^{k} \varphi = \sum_{i=1}^{k} \alpha_{k} p_{k}
\end{equation}
The projection $P^{k}$ increases with $k$:
\begin{equation}
 0 \leq \langle \varphi| P^{k}\varphi\rangle \leq \langle\varphi |P^{k+1}\varphi\rangle \leq \langle\varphi | \varphi\rangle
\end{equation}
\begin{equation}
   0 \leq \langle \varphi| H'P^{k}\varphi\rangle \leq \langle\varphi |H'P^{k+1}\varphi\rangle \leq \langle\varphi |H' \varphi\rangle
\label{Baz_eq2}
\end{equation}
From Eq.~(\ref{Baz_eq2}) we can see that $H'P^{k} \leq H'P^{k+1}
\leq H'$, and we now define intermediate Hamiltonian as
\begin{equation}
H^{k} = H^{0} + H'P^{k}
\end{equation}
It is important to to emphasize that the finite rank operator
$H^{k}$ acts on the functions $p_{1}, p_{2}, ...$ in the same way
as full Hamiltonian $H$:
\begin{equation}
\label{genFRA} H^{k} |i\rangle = H |i\rangle, \quad i = 1,\ldots,
k
\end{equation}
This is the main property of a some finite rank operators which we
will use below.

Following this idea, we construct such an operator in the angular
space of definition of operator (\ref{K_ang}). The Coulomb
potential in hyperspherical variables has the form $V(\rho,\Omega)
= \frac{1}{\rho}f(\Omega)$, where $f(\Omega)$ is the angular part
of potential. We use a finite rank approximation in it. Namely,
the function $f(\Omega)$ is replaced by a finite rank operator:
\begin{equation}
\label{FRAi_op} f(\Omega) \rightarrow {\hat f}^{N} =
\sum_{i,j}^{N} f |\varphi_{i} \rangle d_{ij} \langle \varphi_{j}|f
\end{equation}
Here $\varphi_{j}$ are some auxiliary functions defined in angular
space, $d_{ij} = \langle\varphi_{i}|f|\varphi_{j}\rangle^{-1}$ ---
inverse matrix element.


\section{Formalism}

Here we derive a system of coupled one-dimensional integral
equations using the finite rank approximation. Let us start with
the Schr\"odinger equation:
\begin{equation}
\label{FRAi_1} (H_{0} + V)|\Psi\rangle = E |\Psi\rangle
\end{equation}
Here $H_{0}$ is kinetic energy, $V$ --- interaction potential.
For bound states this equation can be written in integral form using
free Green function:
\begin{equation}
\label{FRAi_2} |\Psi\rangle =
(E - H_{0})^{-1} V |\Psi\rangle =
-G_{E} V |\Psi\rangle
\end{equation}
Let us rewrite it in coordinate representation:
\begin{equation}
\label{FRAi_3} \Psi({\bf R}) = - \int d{\bf R}' G_{E}({\bf R},{\bf
R}') V({\bf R}')\Psi({\bf R'}),
\end{equation}
where ${\bf R} = ({\bf x},{\bf y}) = (\rho, \Omega)$, and use the
Coulombic potential: $V({\bf R}) = \frac{1}{\rho} f(\Omega)$. We
obtain integral equation for the wavefunction $\Psi$ in
hyperspherical coordinates:
\begin{equation}
\label{FRAi_4} \Psi(\rho, \Omega) = - \int {\rho'}^{5} d\rho'
d\Omega' G_{E}(\rho, \rho'; \Omega, \Omega')
\frac{1}{\rho'}f(\Omega') \Psi(\rho', \Omega')
\end{equation}
Using finite rank operator (\ref{FRAi_op}) instead angular part of
potential $f(\Omega)$, we obtain representation for the
wavefunction $\Psi$:
\begin{equation}
\label{FRAi_5} \Psi(\rho, \Omega) = - \sum_{i,j}^{N} \int
{\rho'}^{4} d\rho' d\Omega' G_{E}(\rho, \rho'; \Omega, \Omega')
f(\Omega') \varphi_{i}(\Omega')\, d_{ij}\, C_{j}(\rho'),
\end{equation}
where $C_{j}(\rho') = \int d\Omega'' \varphi_{j}(\Omega'') f(\Omega'')
\Psi(\rho,\Omega'')$ --- new unknown functions.

In order to obtain a system of integral equations for functions
$C_{i}(\rho)$, we use integral operator: $\int d\Omega\,
\varphi_{k}(\Omega)f(\Omega)\ldots$. As a result we obtain:
\begin{equation}
\label{FRAi_5} C_{k}(\rho) = - \sum_{i,j}^{N} \int
d\rho'\,{\rho'}^{4} \int d\Omega\, d\Omega'\,
\varphi_{k}(\Omega)f(\Omega) G_{E}(\rho, \rho'; \Omega, \Omega')
f(\Omega') \varphi_{i}(\Omega')\, d_{ij}\, C_{j}(\rho')
\end{equation}
or
\begin{equation}
\label{int_eq} C_{k}(\rho) = - \sum_{i,j} \int d\rho'\,
M_{ki}(\rho, \rho')d_{ij}\, C_{j}(\rho'),
\end{equation}
\begin{equation}
\label{in_ME}
 M_{ki}(\rho, \rho') = {\rho'}^{4} \int d\Omega\,
d\Omega'\,\varphi_{k}(\Omega)f(\Omega) G_{E}(\rho, \rho'; \Omega,
\Omega') f(\Omega') \varphi_{i}(\Omega')\,
\end{equation}

The Green function $G_{E}({\bf R},{\bf R}')$ has the simplest form in the
momentum representation.
Using the plane wave expansion:
\begin{equation*}
\frac{1}{(2\pi)^3}e^{i{\bf qx}+i{\bf py}} =
\frac{1}{(\kappa\rho)^2}\sum_{KLMl_{1}l_{2}}i^{K}J_{K+2}(\kappa\rho)\,
{\mathcal Y}_{KLM}^{l_{1}l_{2}}(\Omega_{\rho})\,{\mathcal
Y}_{KLM}^{l_{1}l_{2}}(\Omega_{\kappa}),
\end{equation*}
in 
\begin{equation*}
G_{E}({\bf x},{\bf y}) = \iint \frac{d{\bf p} d{\bf q}}{(2\pi)^6}
\exp(i{\bf p}{\bf x}+i{\bf q}{\bf y}) \frac{2m/\hbar^{2}}{p^{2} +
q^{2} + \kappa^{2}},
\end{equation*}
we obtain partial harmonics of free Green function: 
\begin{gather*}
G_{E}^{K}(\rho,\rho') = \iint{\mathcal
Y}_{KLM}^{l_{1}l_{2}}(\Omega)\, G_{E}({\bf R},{\bf R}')\,
{\mathcal Y}_{KLM}^{l_{1}l_{2}}(\Omega_{\kappa})d\Omega\,d\Omega'
=
\\
=  \int_{0}^{\infty}\frac{\kappa
\,d\kappa}{(2\pi)^3}\left(\frac{\rho'}{\rho}\right)^{2}
J_{K+2}(\kappa\rho)\,J_{K+2}(\kappa\rho')\frac{1}{\kappa^2+2mE}=
\\
=  \frac{1}{(2\pi)^3}\left(\frac{\rho'}{\rho}\right)^{2}
  \begin{cases}
  I_{K+2}(\kappa_{0}\rho)K_{K+2}(\kappa_{0}\rho'), \quad 0 \leq \rho \leq \rho'\\
  K_{K+2}(\kappa_{0}\rho)I_{K+2}(\kappa_{0}\rho'), \quad 0 \leq \rho' \leq \rho
  \end{cases}
\end{gather*}
Here $J_{n}(x)$, $I_{n}(x)$ and $K_{n}(x)$ are Bessel function and
modified Bessel functions of first and second kind, respectively, $\kappa^{2} = p^{2}+q^{2}$.
Now we can calculate the kernels of integral
equations~(\ref{FRAi_5}):
\begin{equation}
\label{FRAi_6} M_{ki}(\rho, \rho') = \sum_{KLMl_{1}l_{2}}
G_{E}^{K}(\rho,\rho') \langle\varphi_{k}|f|{\mathcal
Y}_{KLM}^{l_{1}l_{2}}\rangle \langle {\mathcal
Y}_{KLM}^{l_{1}l_{2}}|f|\varphi_{i}\rangle
\end{equation}
We derived a system of coupled one-dimensional integral equations
(\ref{int_eq}). Now one need to calculate the kernel and solve
this system numerically. At this stage of treating the Coulomb
three-body problem the finite rank approximation makes it
sufficiently easier.

\section{Calculation and results}

We constructed the finite rank operator (\ref{FRAi_op}) using
hyperspherical functions. They have been chosen for convenience,
but one can use some other set of linearly independent
functions defined in angular space.

It is important to mention that the representation (\ref{FRAi_5})
for solution of Schr\"odinger equation is not a well known
hyperspherical expansion. One can see it from the definition of
$C_i(\rho)$.

We performed calculations using finite rank operators constructed
on 1, 3 and 6 auxiliary functions. In calculation the kernel
(\ref{FRAi_6}) we should summate an infinite number of terms, but
we stopped at values of the hypermomentum $K$ equal to 6, 10 and
14. In order to solve integral equations, the variables $\rho$ and
$\rho'$ were discretized with 100 mesh points.

We calculated binding energies of the ground state of such
systems: $He$, $H^{-}$, $H_{2}^{+}$, $pp\mu$ and $dd\mu$. Results
of these calculations are presented in Table \ref{tab:1}.

\begin{table}[h]
\caption{Calculated and precise binding energies, eV} \label{tab:1}
\begin{tabular}{c|c|c|cccc}
\hline
 $\quad$           & $E_{pr}, eV$  &  FRA & $K_{max}=6$   & $K_{max}=10$  & $K_{max}=14$ \\
\hline
 $     $           &               &  N=1 &     25.0      &      20.0     &    18.2      \\
 $H^{-}$           &    14.34      &  N=3 &     20.0      &      18.5     &    17.1      \\
 $     $           &               &  N=6 &     18.0      &      16.2     &    15.6      \\
\hline
 $  $              &               &  N=1 &    102        &       98      &     95       \\
 $He$              &    79.0       &  N=3 &     99        &       91      &     89       \\
 $  $              &               &  N=6 &     95        &       87      &     85       \\
\hline
 $         $       &               &  N=1 &      7.3      &       10.0    &     11.0     \\
 $H_{2}^{+}$       &    16.25      &  N=3 &      8.5      &       12.0    &     13.7     \\
 $         $       &               &  N=6 &     10.1      &       13.5    &     15.1     \\
\hline
 $     $           &               &  N=1 &     1050      &      1700     &      1850    \\
 $pp\mu$           &     2782      &  N=3 &     1360      &      2044     &      2101    \\
 $     $           &               &  N=6 &     1690      &      2290     &      2332    \\
\hline
 $     $           &               &  N=1 &     1200      &       1820    &      1990    \\
 $dd\mu$           &     2988      &  N=3 &     1540      &       2072    &      2480    \\
 $     $           &               &  N=6 &     1845      &       2195    &      2654    \\
\hline
\end{tabular}
\end{table}
The precise energies are taken from \cite{Richar_92}.

\section{Conclusion}
\label{concl}
Binding energies of different three-body Coulombic systems were
calculated within a finite rank approximation method. The finite
rank approximation is made in an angular part of potential in
three-body Hamiltonian. This method was tested on some of these
systems earlier in \cite{BelSII_06}.
The results obtained shows it can be useful for solving the Coulombic three-body problem.

One can expect that this method will also be applicable to three charged particles
in continuum, since the approximation (\ref{FRAi_op}) is made at the operator level, but
not in the wavefunction.

\end{document}